\documentclass[reprint,amsmath,amssymb,aps]{revtex4-1}
\usepackage{epsfig}
\usepackage{times}
\usepackage{xcolor}
\usepackage{color}

\begin{document}

\title{When costly migration helps to improve cooperation}

\author{Hsuan-Wei Lee}

\author{Colin Cleveland}
\affiliation{Institute of Sociology, Academia Sinica, Taiwan}

\author{Attila Szolnoki}
\affiliation{Institute of Technical Physics and Materials Science, Centre for Energy Research, P.O. Box 49, H-1525 Budapest, Hungary}
\email{szolnoki.attila@ek-cer.hu}

\begin{abstract}
Motion is a typical reactions among animals and humans trying to reach better conditions in a changing world. This aspect has been studied intensively in social dilemmas where competing players' individual and collective interests are in conflict. Starting from the traditional public goods game model, where players are locally fixed and unconditional cooperators or defectors are present, we introduce two additional strategies through which agents can change their positions of dependence on the local cooperation level. More importantly, these so-called sophisticated players should bear an extra cost to maintain their permanent capacity to evaluate their neighborhood and react accordingly. Hence, four strategies compete, and the most successful one can be imitated by its neighbors. Crucially, the introduction of costly movement has a highly biased consequence on the competing main strategies. In the majority of parameter space, it is harmful to defectors and provides a significantly higher cooperation level when the population is rare. At an intermediate population density, which would be otherwise optimal for a system of immobile players, the presence of mobile actors could be detrimental if the interaction pattern changes slightly, thereby blocking the optimal percolation of information flow. In this parameter space, sophisticated cooperators can also show the co-called Moor effect by first avoiding the harmful vicinity of defectors; they subsequentially transform into an immobile cooperator state. Hence, paradoxically, the additional cost of movement could be advantageous to reach a higher general income, especially for a rare population when subgroups would be isolated otherwise.
\end{abstract}

\maketitle

\textbf{What should we do if we want to avoid being exploited by others or to realize a more supportive environment? A simple answer is to leave our location and search for a better place. This idea was studied intensively in evolutionary game theoretical models where cooperation and defection compete. The problem has been investigated from different angles, where an individual's ambition to leave a detrimental place or search for a more promising location could be the main motivation. Interestingly, the majority of previous models assumed homogeneous populations where everyone is ready to move, ignoring the extra investment required to permanently monitor our neighborhood. In the present research, we explore both aspects and assume that staying or thinking about leaving could be competing options, with explicit consideration to the additional cost of the latter. In the resulting 4-strategy model, we provide insight into the conditions under which costly migration could be beneficial to attain higher general well-being across the entire society.}

\section{Introduction}
\label{intro}

Wars, climate change, and decaying living conditions are the major factors that spur migration, which not only hugely affects the global economy, but also precipitates further change in our hectic world. In 2019, for instance, at least 3.5 \% of the world's population comprised international migrants, according to the UN \cite{UN_20}. Individual mobility, however, is not just a simple reaction to a specific living condition; it could also be the source of a co-evolutionary mechanism of how collective well-being evolves in the framework of human cooperation \cite{perc_bs10,xu_c_pre22,ohdaira_srep22,pi_b_c22,kang_hw_pla21,tao_yw_epl21,couto_njp22,liu_fl_amc22}. Therefore, when we sought to understand the evolution of cooperation in spatially structured populations, considering mobile agents was among our core ideas from the outset \cite{kelly_e92,vainstein_jtb07,sicardi_jtb09}.  

Previous models have focused on different aspects of individual motivations to move. An involved actor's reaction could be risk-driven \cite{chen_xj_pre12b} or environment-driven \cite{xiao_sl_epjb22,park1,park2,park3}, but aspiring to a declared income or expectation regarding a certain payoff level could also be a prime goal of migration \cite{lin_yt_pa11,wu_t_pre12,chen_ys_pa16}. The principal question along this research avenue is how migration influences the evolution of cooperation among players who are involved in a social dilemma where their individual and collective interests are in conflict. More precisely, they can collect a higher payoff individually if they defect, but if they all choose this strategy, then they will earn a smaller income than in the scenario when they cooperate. Interestingly, the conclusions are more subtle in terms of summary in the form of a simple message \cite{cong_r_srep17,park_csf22,li_qr_pa20,chen_w_pa16,xiao_zl_njp20,dhakal_rsos22}. In some cases, the players' movement is proved to be harmful, while in other cases, it clearly becomes a mechanism for supporting cooperation. Additionally, there are examples where the intensity of mobility is a decisive factor in the evolutionary outcome \cite{lutz_pre21,cardinot_njp19}.

Nevertheless, previous model studies typically assumed that a homogeneous population where every player has the capacity to move and improve their individual payoff.  This choice could be the result of a decision given that moving could be a costly skill. More precisely, those players who are willing to move should constantly watch and evaluate their neighborhood, as whether to move will depend on the result of this analysis. Undoubtedly, these efforts require an extra investment from the mentioned players, which should be considered an extra cost. Perhaps it is worth mentioning that one previous study examined costly migration \cite{liu_yk_csf12}, but the conclusions were controversial. To solve the puzzle and provide a more comprehensive conclusion in our research, we assume that the population could be heterogeneous. More precisely, a player can decide whether she wants to invest extra efforts to explore the neighborhood and possibly change residence, or whether, alternatively, she prefers to save in terms of the payoff by not moving and using unconditional cooperator or defector strategies. Premised on this, we constructed a 4-strategy model where both approaches are available, allowing us to explore the profitability of each attitude at specific parameter values, namely, average population density and the strength of the social dilemma. In the next section, we define our extended public goods game (PGG) and introduce the key parameters. Sec.~\ref{result} highlights our main observations. Sec.~\ref{sum} concludes with a summary of the results and a discussion about their implications.

\section{Model}
\label{def}

We start from a traditional spatial PGG, where players are distributed on a square lattice with a linear size of $L$, and periodic boundary conditions are applied \cite{sigmund_10,perc_jrsi13}. Some of the lattice points are occupied by a player while others are not; hence, altogether, $0< \rho \le 1$ portion of the available $L \times L$ positions are filled. According to the standard protocol, an actor, who could be an unconditional cooperator ($C$) or an unconditional defector ($D$), plays a PGG with the nearest available neighbors in which a cooperator player invests the amount of $c=1$ in the common pool, while a defector player refrains from doing so. We then increase the accumulated contributions by an $r$ enhancement factor and distribute the resulting public goods equally amongst all group members regardless of their strategies. Of course, players are also involved in the games their neighbors organize.

We therefore extend the model by introducing two new strategies that behave similarly to the original strategies in the game but entail an extra skill related to moving. More precisely, it is well-known that a cooperation level that is too low is detrimental to everyone. Therefore, the new group of players, called sophisticated players, are capable of evaluating their local neighborhood and leaving it if it is harmful. In particular, they will leave their original positions if the average cooperation level is less than 50\% among the existing neighbors. Moreover, if there is no other actor in the nearest neighborhood because all four neighboring lattice sites are empty, the actor will leave the site, too. In this way, we introduce sophisticated cooperator ($SC$) and sophisticated defector ($SD$) strategies. These players always explore their environment, and if they are dissatisfied, they move to a randomly chosen neighboring empty place and make a trial strategy change in the new position. Importantly, sophisticated players bear an extra cost $f$ because they need to constantly watch and evaluate their neighborhood. One may argue that the introduction of mobile players will definitely support cooperation because cooperators may have a chance to escape from a situation where they are being exploited. On the other hand, mobile defectors also have the opportunity to search for remote players who could be new prey given their selfish attitude. Hence, the question is far from trivial. Interestingly, however, as we will show, this extra cost has a biased consequence that impacts the competing strategies. 

The strategy change is based on the widely used imitation process, where the probability of a player $x$ adopting the strategy of a neighboring player $y$ is based on the payoff difference between these competitors:
\begin{equation}
W(s_x,s_y)=\frac{1}{1+\exp[(\Pi_{x}-\Pi_{y})/K)]},
\label{fermi}
\end{equation}
where parameter $K$ quantifies the uncertainty of strategy imitation. To attain results that are comparable with those previously obtained in the traditional model, we fixed $K=0.1$.

In sum, at the beginning of the Monte Carlo (MC) simulation, we first randomly distribute $N= \rho \times L \times L$ players on a square lattice where $\rho$, as a key parameter, determines the average population density. The players randomly choose a strategy from the available set $(C, D, SC, SD)$. During an elementary step, we randomly choose a player $i$. If the strategy that player $i$ selects is $SC$ or $SD$, which means we have a sophisticated player, then player $i$ explores her neighborhood and determines the average cooperation level among the neighbors. If it is below $1/2$, then the player moves to a randomly chosen empty neighboring place. In the next step, player $i$, either in the original position or in the new place, selects a neighboring player $j$ who could be a potential model player for strategy adoption. After we calculate the payoff values for both involved players according to the PGG protocol, it is reduced by cost $f$ for a sophisticated player. Next, player $i$ imitates player $j$ with the probability defined by Eq.~\ref{fermi}. Evidently, traditional players are only involved in the strategy updating step at their original positions. A full MC step is established if we repeat the above-described procedure $N$ times.  The relaxation time to reach the stationary state depends not only on the applied system size but also on the population density. In each case, we monitored the evolution of strategies and measured the proportions of strategies when they reached the stationary values. The typical relaxation time was about 10,000 MC steps. In this research, we typically applied a linear system size of $L=400$, but we also checked other sizes to exclude finite-size effects. To obtain reliable values for the cooperation level or for the proportions of strategies, we have typically averaged the results of 10-50 independent runs.

\section{Results}
\label{result}

Let us begin by presenting our key observation about the consequence of mobile players' presence in addition to traditional immobile actors. This can be done effectively if we compare the general cooperation level for the original 2-strategy model with that of the extended 4-strategy case defined in the previous section. As we noted, our main control parameters are the $r$ synergy factor of PGG, which characterizes the strength of the dilemma, and $\rho$ general density, which indicates the population's degree of rarity or crowdedness.

\begin{figure}
\begin{center}
\includegraphics[width=8.5cm]{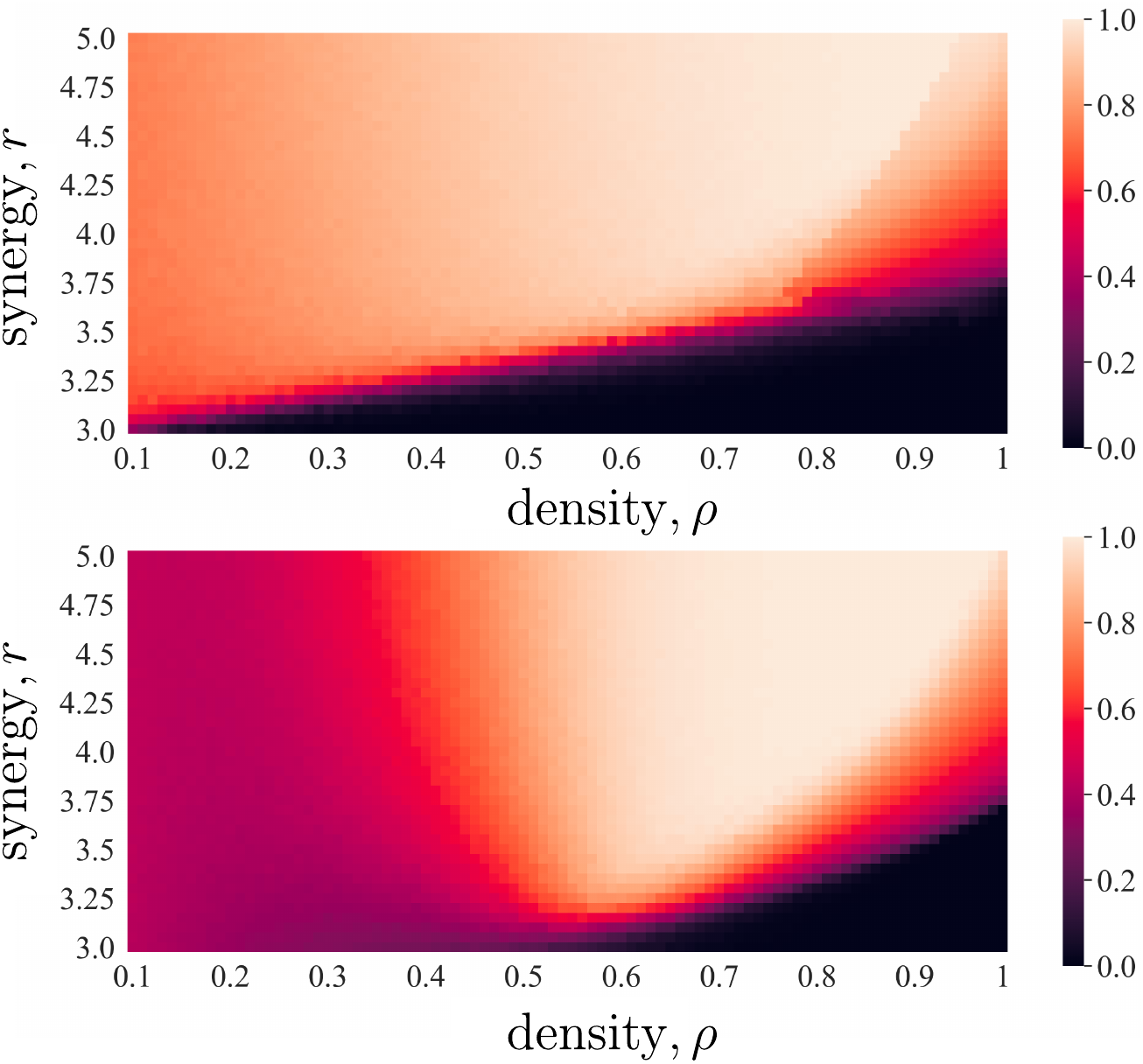}
\caption{Cooperation level on the enhancement factor--density parameter plane for two models. The upper panel shows the result of the 2-strategy model where unconditional cooperators ($C$) and defectors ($D$) are present and players have no moving capacity. The bottom panel shows the cooperation level for our present 4-strategy model, which represents our extension of the original model with the addition of two new strategies that entail cooperation ($SC$) or defection ($SD$), and players are capable of leaving their positions if they find them detrimental. This skill requires extra effort, in the form of an extra fee, which is $f=0.08$ in the plotted case. Evidently, the sum of the proportions of $C+SC$ fractions indicates the cooperation level in the latter case.} \label{heat}
\end{center}
\end{figure}

A comparison of the results obtained in the mentioned cases is presented in Fig.~\ref{heat}. The upper panel shows the traditional model, where we can see that there is an optimal density that provides the highest cooperation level, especially at modest $r$ values. This result confirms a previous finding that at a fixed dilemma strength, we can reach a higher cooperation level if the population is less crowded \cite{wang_z_pre12b}. More precisely, the highest cooperation level can be reached in the vicinity of the percolation threshold, which provides a successful combination of external conditions: The population is sufficiently rare to avoid providing easy prey for defectors, but on the other hand, information about cooperators' collective success can spread throughout the whole system \cite{wang_z_srep12}. The latter condition is not fulfilled if the population is rare. In this case, it is almost a random situation in which strategy wins on a small island and the local evolutionary outcome depends largely on the initial distribution of strategies. Therefore the final and practically frozen state is a mixture of pure $C$ and pure $D$ islands. This consequence can be seen nicely in the upper panel of Fig.~\ref{heat}, where the cooperation level is around 0.5, almost independently of the dilemma strength if $\rho$ is sufficiently small.

This situation changes significantly if mobile strategies are also introduced. In this case, a significantly higher cooperation level is also reached in a rare population. One might say that the information spreads due to mobile agents, but the explanation is not as simple because defectors can also move. The deeper answer is based on the highly biased consequence of costly movement, which is harmful to defectors but can offer cooperators extra help. We will elaborate on this argument later. Here, it is worth noting that the applied extra fee $f$ is sufficiently high to represent the enhanced investment for sophisticated players, but it is not too large to make their payoff values unfeasible.

\begin{figure}[h!]
\begin{center}
\includegraphics[width=8.0cm]{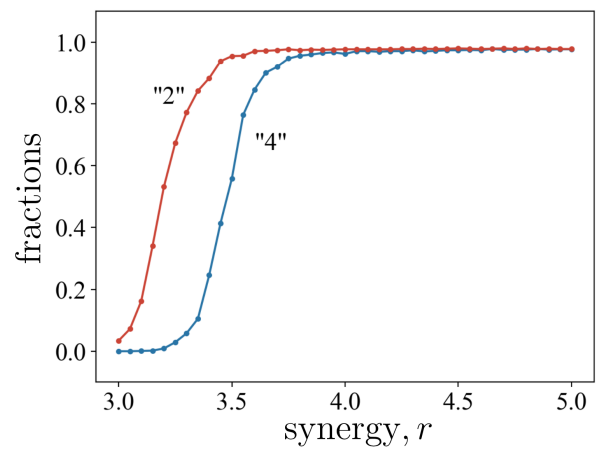}
\caption{Cooperation level dependent on the enhancement factor at $\rho=0.65$ for the traditional 2-strategy model (red) and the extended 4-strategy model (blue), with $f=0.08$, as indicated. Around the optimal density, which is close to the percolation threshold, mobile actors' presence could be harmful when the synergy is low.} \label{bad-4}
\end{center}
\end{figure}

If we carefully compare the mentioned panels, we can see that a better outcome can be reached in the traditional 2-strategy model around the previously mentioned optimal $\rho \approx 0.6$ density when $r$ is small; hence, the dilemma strength is more critical. For a quantitative comparison, in Fig.~\ref{bad-4}, we simultaneously present both models' cooperation levels dependent on the synergy factor at a fixed value of $\rho=0.65$. Evidently, in the 4-strategy model, the cooperation level is calculated as the sum of the proportions of the $C$ and $SC$ strategies. The figure shows that a significantly higher cooperation level can be reached in the traditional model, where players cannot move, and the difference only vanishes for higher $r$ values. This effect can be understood if we consider how movements happen. If a player is satisfied because the cooperation level is sufficiently high, then even mobile players will maintain their positions and move only in the alternative case. 

This so-called environment-driven movement can cause slight changes in the interaction graph, resulting in more aggregated colonies. The aggregation of players can be measured directly if we calculate the probability of finding nearest-neighbor players around a player. In this way, two-point probability $p(1,1)$ can indicate the likelihood of finding the nearest neighbor next to our focal player. For example, if we consider the two-strategy model at the mentioned density of $\rho=0.65$, then the supposed random distribution of players results in an approximate value of $p(1,1)=0.42$ for the mentioned pair probability. This value is significantly larger for the 4-strategy model, where players may move and approach each other. More specifically, if we use the same $\rho$ value at $r=3.4$ and $f=0.08$, then the mentioned probability increases to 0.60. In other words, the population becomes more aggregated, but it also means that the comprehensive percolation is in jeopardy because we are close to the critical $\rho$ value. In this way, mobility may threaten one of the conditions we previously identified as necessary to reach the highest cooperation level. Of course, if the synergy factor is sufficiently high, then we do not need proper percolation because the cooperator can also win locally; hence, the detrimental consequence of the player's movement becomes irrelevant, and the cooperation level is restored to reach the value produced by the simpler model with fixed players. To summarize our observations, there is a small area on the parameter plane at intermediate $\rho$ values and modest $r$ values when the traditional model provides a better outcome, while for the majority of parameter pairs, the 4-strategy model undoubtedly offers a higher cooperation level.

\begin{figure}
\begin{center}
\includegraphics[width=8.0cm]{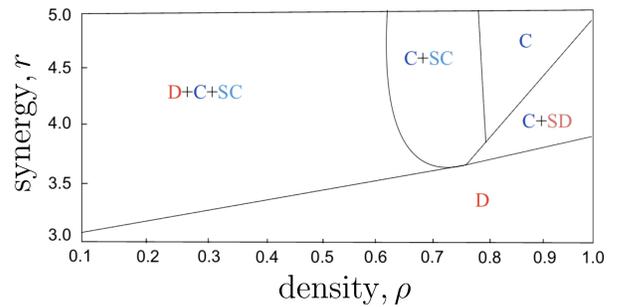}
\caption{Phase diagram of the 4-strategy model on the $r-\rho$ plane obtained for an additional cost $f=0.08$. If the synergy factor is sufficiently low, then only fixed defectors survive, independent of the average population concentration. Sophisticated defectors can only survive in a limited parameter area if the dilemma is mild and the system is crowded. Sophisticated cooperators, however, are viable in a large area, especially when mobility has an enhanced role in a rare population.} \label{phd}
\end{center}
\end{figure}

In the following paragraphs, we provide deeper insight into the consequences of adopting new strategies. In Fig.~\ref{phd}, we present a phase diagram on the $r-\rho$ parameter plane, where the same extra cost value for sophisticated players is used as in Fig.~\ref{heat}. First, we would like to stress that only traditional defectors survive if the synergy factor is sufficiently low, independent of the average density $\rho$. Evidently, if the social dilemma is strong, then cooperators cannot gain the necessary income to compensate for defectors' exploitation. Consequently, only those who adopted $D$ and $SD$ strategies survive at the intermediate evolutionary stage. In this case, however, the extra cost of the $SD$ strategy forms a disadvantage comparing with simple $D$ players, who will eventually dominate the whole system.

Interestingly, the advantage of the $D$ over the $SD$ strategy remains even if the extra cost $f$ is zero. In the latter case, one might expect that the capacity to move should provide some benefit to adopters of the $SD$ strategy, but the opposite is true. The reason is as follows. The advantage of $SD$ to move also makes accumulation harder; it is therefore easier for an $SD$ player to meet a $D$ player in a new position. On the contrary, if we select a $D$ player, who cannot escape from nearby players for strategy update, then it is more likely that the $D$ player will choose another $D$ neighbor as a model player, in which case nothing happens, and the portion of $D$ players remains unchanged. When we select an $SD$ player during the elementary process, then she may move first, but she may sequentially encounter a $D$ player. Indeed, there is no difference between their payoff values, but Eq.~\ref{fermi} indicates a 0.5 probability of adopting strategy $D$, which can decrease the portion of $SD$. This asymmetry will gradually lead to a fully $D$ destination.

Returning to the phase diagram, it highlights that the $SD$ strategy is only viable in a very limited parameter area when the dilemma strength is modest and the population is practically crowded. At these $r$ values, the coexistence of cooperator and defector strategies is expected in a spatial system \cite{szolnoki_pre09c}; therefore, $SD$ players can enjoy the proximity of supporting cooperators that can generously cover their extra fees. On the other hand, their skill of leaving a detrimental environment makes them more competitive relative to the pure $D$ strategy. In other words, our preliminary expectation about the advantage of mobile players is justified in the mentioned case. Importantly, however, it is only valid for the $SD$ strategy, not the $SC$ strategy. Players who adopt the latter strategy simply have no chance of escaping defectors because of the high general density, making it evident that the extra cost puts them at a disadvantage. These mechanisms explain that $C+SD$ forms a stable solution here.

When the system is less crowded, another aspect of the actor's mobility can be detected. In a rare population, cooperators may have a good chance of escaping defectors; hence, $SC$ strategy becomes viable. Interestingly, intermediate concentration does not provide an opportunity for the $SD$ strategy. Indeed, these $SD$ adopters can escape from nearby players, but they can no longer utilize cooperator strategies, because the latter type of player can form a compact island, which gives them highly competitive payoff values. As we stressed above, intermediate density was already a perfect condition in the 2-strategy model, and this conclusion remains intact in the extended model. One may claim that $C$ and $SC$ are still unequal since the latter should bear the extra cost $f$. However, they remain separate because $SC$ players can form ``satisfied'' groups; hence, their coexistence is a mixture of the $C$ and $SC$ domains.

Finally, if the population is very rare, then immobile defectors once again have a chance at survival. In this case, they win their local battle, but they are stuck afterward. Other domains invaded by $C$ or $SC$ players are also satisfied; they remain still, and the information about more successful strategies does not reach $D$ players. This explains why we can observe the $D+C+SC$ solution here.

\begin{figure}
\begin{center}
\includegraphics[width=8.0cm]{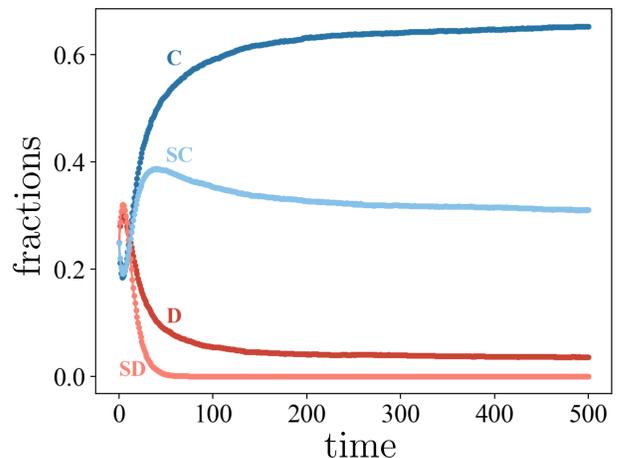}
\caption{The early evolution of the proportions of strategies in the extended model at $r=4$, $\rho=0.65$, and $f=0.08$. We can detect three different stages in the evolutionary process, which are discussed thoroughly in the main text. While the early growth of $C$ and $SC$ strategies after an initial decline is a sign of how network reciprocity works, the subsequent decline of the $SC$ strategy is a consequence of the so-called Moor effect. Finally, though this is not shown here, all defector strategies die out, and only cooperator strategies survive.} \label{moor}
\end{center}
\end{figure}

Before discussing these parameter regions, let us return to the case of intermediate $\rho$ values; we can observe an interesting process if we monitor the proportion of strategies during the evolution.  Typical time evolution is shown in Fig.~\ref{moor}, where we plotted how the proportions of strategies change if we launch the evolution from a random initial state at $r=4$, $\rho=0.65$, where sophisticated players bear an extra cost of $f=0.08$. At the very beginning, for a short initial period, defector players enjoy a random local neighborhood; hence, cooperators are decaying. The survivors then form cooperative clusters that provide them with competitive payoffs; hence, their numbers start growing. This ``first down, later up'' dynamic is the typical trademark of network reciprocity reported earlier in several cases \cite{perc_pre08b,szolnoki_epjb09,wei_x_epjb21,shen_y_pla22,liang_rh_pre22}. Another consequence of the emergence of cooperative clusters is the rapid extinction of the $SD$ strategy. Even if they can escape nearby players, during their migration, they may encounter a phalanx of cooperators that can easily defeat them. Of course, this effect is not found among pure $D$ players, who cannot move; hence, they are not threatened by a more successful strategy. Accordingly, their proportion declines at a significantly slower rate due to the strategy invasion process. The portion of sophisticated cooperators, however, starts declining again after reaching the maximum level. This phenomenon can be explained as follows. First, it is advantageous for cooperators to be mobile because they can avoid intensive exploitation at the hands of defectors. This is why their minimal proportion in the first stage is a bit higher than the minimum level for the $C$ strategy. Hence, the presence of the $SC$ strategy helps to fight defection more effectively. However, later, when defectors are eradicated, the vicinity of the simple $C$ strategy becomes dangerous to them because the negative consequence of the additional cost $f$ lowers their payoff. This process resembles the so-called Moor effect \cite{szolnoki_pre11b}. When sophisticated cooperators have performed their duties (and weakened defectors), they may go (for the benefit of simple cooperators).

\begin{figure}
\begin{center}
\includegraphics[width=8.0cm]{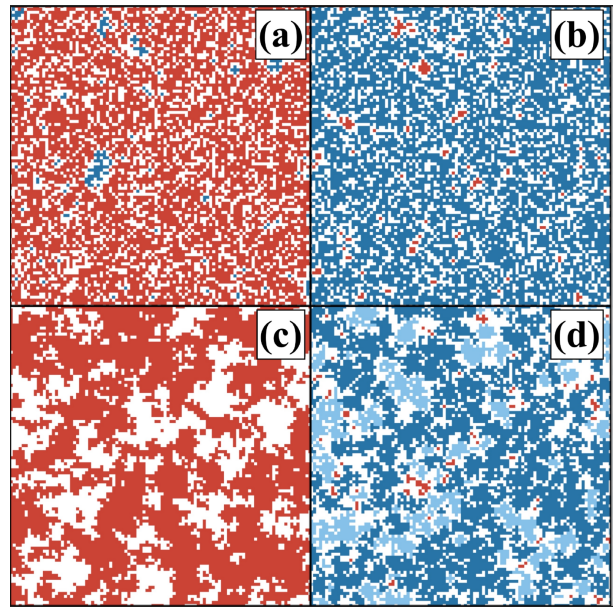}
\caption{Representative strategy distributions obtained for a density of $\rho=0.65$ at $r=3$ (left column) and at $r=5$ (right column). Panels~(a) and (b) show the 2-strategy model, while panels (c) and (d) depict the 4-strategy case. The panels only show a $100 \times 100$ section of a larger system. Dark (light) blue represents pure (sophisticated) cooperators, while dark red shows pure defectors. Note that sophisticated defectors cannot survive here.} \label{pattern_65}
\end{center}
\end{figure}

In Fig.~\ref{pattern_65}, we present some characteristic patterns in this parameter area to summarize the differences between the traditional and the extended models. The first row shows the stationary strategy distribution in the 2-strategy model while the bottom row depicts the results of the 4-strategy case. Furthermore, the patterns shown in the first column were obtained at $r=3$, which represents harsh external conditions, while the second column illustrates observations for a more friendly case at $r=5$. The difference between the two is striking. While the players' spatial distribution is random in panels (a) and (b) due to the model's similarity to the traditional 2-strategy model, the players are more aggregated when individual mobility is allowed. This is especially true for a small $r$, as shown in panel~(c), because in this case, nobody is satisfied in the initial stage; hence, sophisticated players move actively. However, they are prone to easy capture by defectors because of the low $r$ value. In the large $r$ case, the population is still more aggregated than a random distribution; hence, the pattern is below the percolation threshold, though the total number of players would dictate the percolation in a uniformly distributed system. This explains why $SC$ players can survive: because they can form isolated domains comprising pure $C$ players who would otherwise beat them. For a similar reason, some isolated $D$ spots remain, but they represent a negligible minority; hence, the system closely approaches a full cooperator state.  

\begin{figure}[h!]
\begin{center}
\includegraphics[width=8.0cm]{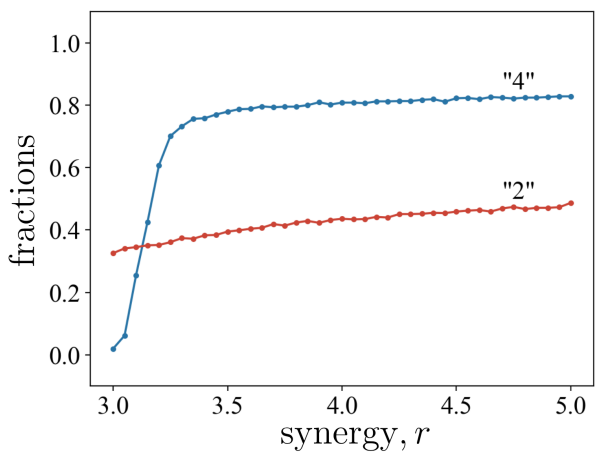}
\caption{Cooperation level dependent on the enhancement factor at $\rho=0.25$ for the traditional 2-strategy model (red) and the extended 4-strategy model (blue), with $f=0.08$, as indicated. In this parameter region, the presence of sophisticated players has a clear positive impact on public cooperation.} \label{good-4}
\end{center}
\end{figure}

Next, we discuss the low $\rho$ region when the population is rare and smaller communities become partly separated. In this case, the introduction of costly mobile strategies has a clear positive impact regarding improving general cooperation. This can be seen in Fig.~\ref{good-4}, where we compared the average cooperation levels dependent on the synergy factor at $\rho=0.25$. First, we briefly discuss the gentle increasing red slope of the 2-strategy model. As we mentioned at the beginning of Sec.~\ref{result}, here, the key parameter that characterizes the dilemma strength only has second-order importance. There are small communities where the evolutionary outcome is sensitive to the initial distribution of strategies. Therefore either cooperators or defectors can win locally, which provides an average cooperation level close to the random value. Of course, the value of the synergy factor plays a certain role; hence, the cooperation level increases slightly as we increase $r$. However, the $r$-dependence is unusually weak, which is the consequence of poor communication between smaller communes. Of course, it also means that cooperators may survive at very small $r$ values, but practically, it is an artificial effect, and it is divorced from the social dilemma.

This situation changes dramatically when we introduce mobile players. In this case, a significantly higher cooperation level can be reached at a low $r$ value because mobile players can spread the good news about the benefits of cooperation. Of course, $SD$ players can also move, but they are unprotected, as we argued earlier. If we compare the blue line with the results shown in Fig.~\ref{bad-4}, we see that full cooperation cannot be reached in a rare population, even if we have mobile actors. This is because some tiny domains are occupied by pure $D$ players. Even if they are unsatisfied, they cannot change their status. Other domains formed by $C$ players are also immobile, or, if they are $SC$ players, then they are satisfied in the company of nearby neighbors. Therefore, they are not motivated to move; hence, the poor performance of defector spots remains hidden, which noticeably lowers the general cooperation level. Interestingly, this mechanism cannot work at an intermediate density because defector domains cannot be completely isolated. It is also worth noting that the above-mentioned Moor effect regarding the time evolution of the $SC$ strategy is less visible at small densities, the explanation for which is similar to that for the survival of fixed defectors. Specifically, isolated $SC$ actors may reach a state of satisfaction in the vicinity of similar players; hence, they do not move, and can keep their original strategy because their additional cost no longer has any relevance.

\begin{figure}
\begin{center}
\includegraphics[width=8.0cm]{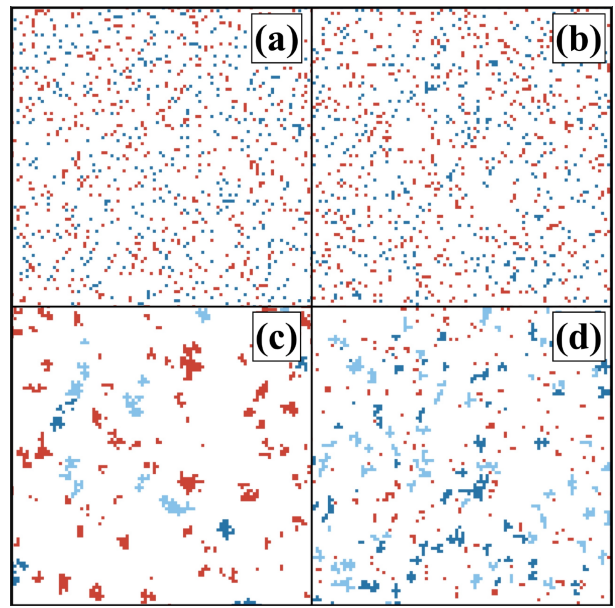}
\caption{Representative strategy distributions obtained for a density of $\rho=0.1$ at $r=3$ (left column) and at $r=5$ (right column). Panels (a) and (b) show the 2-strategy model, while panels (c) and (d) depict the 4-strategy case. The panels only show a $100 \times 100$ section of a larger system. The color code is the same as in Fig.~\ref{pattern_65}.} \label{pattern_10}
\end{center}
\end{figure}

Next, we support our arguments by presenting some representative patterns obtained at $\rho=0.1$. In Fig.~\ref{pattern_10} we follow the same logic as in Fig.~\ref{pattern_65}. As the top row highlights, we have small isolated homogeneous domains in the final state, where either $C$ or $D$ wins. If we compare the columns, there is no significant detectable difference between the configurations shown in panels (a) and (b). Here, the $r$ value has no crucial role because the final local outcome is highly sensitive to the initial distribution of strategies. 

Turning to the 4-strategy case, $SD$ cannot survive; hence, we did not detect players represented by light red. Compared to the top row, the aggregation of players is obvious in panels (c) and (d) as a result of sophisticated players' migration. Due to the small $r$ value, a small island should be homogeneous. Here, either cooperation or defection should prevail locally. Notably, both $C$ and $SC$ strategies can survive. Adopters of the latter may reach a state of satisfaction in the vicinity of similar actors and therefore refrain from any further movement. For the high $r$ case, the main difference is that defectors cannot build large domains it is no longer as easy for them to emerge victorious from the initial battle. This mechanism results in a significantly higher general cooperation level. However, some isolated red spots remain, which explains why we cannot realize a full cooperator state here. The latter would require an effective information percolation throughout the whole system.

Finally, we briefly discuss the role of the additional cost parameter $f$, which is a crucial element of our model compared to the large number of other models previously used to understand the impact of migration on the original social dilemma. Evidently, if this cost is too high, then sophisticated strategies become less viral, and we practically revert to the original 2-strategy model. Interestingly, however, the behavior of the $SD$ strategy is almost independent of the value of $f$, and the strategy only survives in a very limited parameter area when $\rho$ is sufficiently high and $r$ is in an intermediate region. From this aspect, the presented behaviors obtained at $f=0.08$ faithfully represent what we can expect of other moderate $f$ values.

\section{Conclusion}
\label{sum}

It is perhaps unnecessary to emphasize the importance of exploring the possible impacts of migration on the outcomes of social dilemmas because individual mobility may generally result in subtle collective behavior among system members \cite{vicsek_pr12}. Therefore, the last decade has seen intensive research activity along this path \cite{zheng_jj_pa22,flores_jtb21,quan_j_pa21,deng_ys_epjb22,lee_hw_amc22,liu_jz_epjb21,lv_amc22,quan_j_csf21,zheng_lp_pa21,li_xy_epjb21,fu_mj_pa21,zhu_pc_epjb21,yang_lh_csf21,li_k_csf21,lee_hw_pa21,quan_j_jsm20,fu_mj_pa19}. In this research, our principal motivation was twofold. First, we wanted to ascertain the impact if players have the liberty to decide whether they want to ponder the option of moving. This aspect can be modeled using two significantly different families of strategies. While the first one focuses on how to behave in the original social dilemma, the other entails permanent observation of the neighborhood, and individuals may move if their neighbors' behavior is unsatisfactory. In other words, it could also be the source of a dilemma regarding whether we want to ponder with the option of moving. If the answer is yes, then we will pay the price for extra information and for maintaining the capacity to change our position. Alternatively, we can consider the cost of proper movement into this parameter. Notably, the majority of previous work has ignored this aspect. 

It is also worth mentioning that we introduced the so-called sophisticated players as strategy-neutral, because both cooperators and defectors may choose this option. Therefore, its impact for the cooperation level is far from trivial. Our main observation was that the introduction of an extra cost has a highly biased effect on the competition among the original strategies. In particular, the sophisticated cooperator strategy is viral and can help to elevate the cooperation level in the majority of the available area within the parameter space. This is especially true in a population where the average density is low and information exchange would be difficult otherwise. There is a very limited parameter space, with an intermediate concentration and a low synergy factor, where sophisticated players' presence is detrimental. In this narrow parameter region, the fixed players would have formed an optimal interaction pattern, which provides a high cooperation level. In the mentioned case, the population is sufficiently rare; hence, defectors have little opportunity to exploit neighboring cooperators. However, because we are beyond the percolation threshold, the positive impact of cooperation can spread throughout the whole system. Allowing some players to move damages the latter condition. Consequently, players aggregate more intensively, and the percolation is broken.

The asymmetric impact of the extra cost on strategies is also evidenced by the fact that sophisticated defectors can only survive in a very limited parameter area in coexistence with pure cooperators. In this section of the parameter plane, the population is essentially crowded; hence, individual mobility only has second-order importance. Furthermore, the applied synergy factor would also allow the coexistence of pure strategies. 

In sum, our model is an initial trial to investigate more realistic models where all aspects of individual mobility are considered, allowing us to accurately estimate the impact of this factor on fundamental social dilemmas. Extensions to other evolutionary games and alternative interaction graphs would increase our understanding of the original question. 

\section*{Acknowledgments}

The research reported was supported by the Ministry of Science and Technology of the Republic of China (Taiwan), under grant No. 109-2410-H-001-006-MY3.

\section*{Author declarations}

The authors declare no competing interests.

\section*{DATA AVAILABILITY}

Data sharing is not applicable to this article as no new data were created or analyzed in this study.

\end{document}